\documentclass[10pt,a4paper, final]{IEEEtran}
\usepackage[english]{babel}
\usepackage[T1]{fontenc}
\usepackage{amsfonts,amssymb,amsthm,mathrsfs}
\newtheorem{remark}{Remark}
\usepackage{array}
\usepackage{graphicx}
\usepackage[caption=false,font=footnotesize]{subfig}
\usepackage{cite}
\usepackage{algorithm}
\usepackage{algpseudocode}
\usepackage{balance}
\usepackage{tikz}
\usetikzlibrary{shapes}
\usepackage{lscape}
\usepackage{mathtools}
\usepackage{comment}
\usepackage{verbatim}
\usepackage{xfrac}
\usepackage{wrapfig} 
\usepackage{comment}
\usepackage{booktabs}
\usepackage{color}
\usepackage{epstopdf}
\usepackage{float}
\usepackage[T1]{fontenc}
\usepackage{multicol}
\usepackage{bm}
\usepackage{cite}
\usepackage{soul} 
\usepackage{url}
\usepackage{hyperref} 
\usepackage[most]{tcolorbox}

\usepackage{tikz}
\definecolor{gold}{rgb}{0.85,.66,0}

\definecolor{crimson}{rgb}{0.66, 0.38, 0.44}

\renewcommand{\mathbf}[1]{\boldsymbol{{\rm#1}}}
\renewcommand{\mathbf}[1]{\boldsymbol{{\rm#1}}}

\begin{document}
\title{Hybrid-Field Channel Estimation for XL-MIMO Systems: Dictionary-based Sparse Signal Recovery}
\author{{David William Marques Guerra}, 
{Taufik Abr\~ao}
\thanks{This work was supported in part by the CAPES (Financial Code 001) and the National Council for Scientific and Technological Development (CNPq) of Brazil under Grant  310681/2019-7.}
\thanks{D. W. M. Guerra is with the Department of Electronics and Systems of the Federal University of Pernambuco (UFPE). Recife 50740-550, PE - Brazil. Email: \url{david.guerra@ufpe.br}.}
\thanks{T. Abrão is with the Department of Electrical Engineering. The State University of Londrina.  Po.Box 10.011; Londrina 86057-970, PR - Brazil. Email: \url{taufik@uel.br}.}
\vspace{-.2mm}
}

\maketitle
\newpage
\color{black}
\begin{abstract}
Extremely large-scale multiple-input multiple-output (XL-MIMO) systems are a key technology for future wireless networks, but the large array aperture naturally creates a hybrid-field (HF) propagation regime in which far-field (FF) planar-wave and near-field (NF) spherical-wave components coexist. This work considers the problem of HF channel estimation (CE) and introduces a unified model that superimposes FF and NF contributions according to the Rayleigh distance boundary. By exploiting the inherent sparsity of the channel in the angular and polar domains, we formulate the estimation task as a sparse recovery problem. Unlike conventional approaches that require prior knowledge of the channel sparsity level, the proposed method operates without requiring knowledge of the sparsity level $L$ and the NF$/$FF ratio $\gamma$, which are used only for synthetic channel generation in simulations. The channel estimator determines the number of paths adaptively through a residual-based stopping rule. A combined FF/NF dictionary is employed to initialize the support, and each selected atom undergoes continuous parameter refinement to mitigate grid mismatch. Simulation results demonstrate that the proposed estimator achieves accurate HF channel reconstruction under both line-of-sight (LoS) and non-line-of-sight (NLoS) conditions, offering a practical and computationally efficient solution for XL-MIMO systems.
\end{abstract}
\begin{IEEEkeywords}
Extremely Large-Scale MIMO (XL-MIMO); Channel State Information (CSI); {Channel estimation (CE)}; hybrid-field (HF) wave propagation; near-field (NF) spherical wave model; far-field (FF)
planar wave model.
\end{IEEEkeywords}

\IEEEpeerreviewmaketitle

\section{Introduction}\label{sec:intro}
Extremely large-scale multiple-input multiple-output (XL-MIMO) is a cornerstone technology for future sixth-generation (6G) wireless networks, promising unprecedented gains in spectral efficiency (SE) and spatial resolution (SR) \cite{An2024}. However, the massive antenna arrays deployed in XL-MIMO systems introduce a unique propagation environment where the traditional {far-field (FF)} plane wave assumption no longer holds for all scatterers \cite{Tian2025}. This gives rise to a hybrid-field (HF) communication scenario in which both {near-field (NF)} spherical wave propagation and FF planar wave propagation coexist. Accurate channel state information (CSI) is critical to realize the full potential of XL-MIMO, but the {NF} nature of the channel presents significant challenges for estimation \cite{Lu2024}.

{In the context of XL-MIMO channel estimation (CE) under HF conditions, some scatterers lie in the FF region of the base station (BS) while others reside in its NF, yielding channels that comprise a mixture of planar and spherical wave components \cite{Zeng2025}. To capture this effect, and in contrast to the conventional FF {\it orthogonal matching pursuit} (OMP) estimator of \cite{Huang2019} (which uses only an FF dictionary and computes gains via pseudo-inversion), \cite{Wei2022} introduced an HF model with an adjustable proportion parameter. This enables separate sparse representations of FF components in the angular domain and NF components in the polar domain. Based on this model, \cite{Wei2022} proposed an HF-OMP procedure that sequentially estimates FF paths and then NF paths using the corresponding dictionaries. Variants such as support detection OMP (SD-OMP) \cite{Hu2023} refine the FF/NF support by evaluating small clusters of neighboring atoms. Since these methods require prior knowledge of the NF/FF ratio, \cite{Yang2023} addressed this limitation by performing an exhaustive search over all possible NF/FF splits ($L_F{+}L_N=L$), reconstructing the channel for each candidate configuration and selecting the one yielding the smallest residual.}

\vspace{.5mm}
\noindent{\noindent{\bf Related work}: Another class of HF/NF estimators relies on sparse gradient projection (SGP) and gridless refinement. The method in \cite{Cui2022} targets NF estimation by combining a polar-domain simultaneous OMP (SOMP) stage with a simultaneous iterative gridless weighted (SIGW) refinement using gradient updates and Armijo line search. More recently, \cite{Lei2024} extended this framework to the HF setting: one variant assumes the NF/FF ratio is known (SGP-with-$\gamma$), refining NF parameters directly using SIGW; another variant removes this assumption by integrating the exhaustive NF/FF search of \cite{Yang2023} with the SIGW refinement of \cite{Cui2022}. Despite their improved accuracy, all these approaches require prior knowledge of the total sparsity level $L$ (and, in most cases, of the NF/FF ratio), which limits their practical applicability in XL-MIMO systems.}

\vspace{.2mm}
\noindent{\noindent{\bf Contribution
}: The main contribution of this work lies in a simplified and computationally efficient formulation for HF XL-MIMO channel refinement. Unlike existing HF and NF methods \cite{Wei2022, Hu2023, Yang2023, Cui2022, Lei2024}, which rely on matrix-based operations such as pseudo-inversion or multi-dimensional gradient steps, the proposed approach derives a closed-form expression for the path gain update, in which thanks to the single-column refinement structure, the gain associated with each newly selected atom is obtained from a {\em scalar} least-squares (LS) solution. Likewise, the refinement of the NF/FF structural parameters relies only on scalar gradients (rather than matrix gradients as in \cite{Cui2022, Lei2024}) leading to a substantially lighter update rule. This yields a much simpler analytical structure and a significant reduction in complexity per iteration. Furthermore, in contrast to all prior HF estimators \cite{Huang2019, Wei2022, Hu2023, Yang2023, Lei2024, Cui2022}, our method does not require prior knowledge of the channel sparsity level $L$ nor of the NF/FF path split $\gamma$. The only required parameter is the noise variance, which can be easily estimated in practice, making the proposed estimator substantially more robust and deployment friendly. Numerical results confirm that the proposed HF CE, called $\epsilon$-OMP–{\textit{simplified sparse iterative gridless weighting} (SSIGW)}, achieves a superior performance–complexity trade-off when compared with eight state-of-the-art benchmarks, highlighting both its accuracy and efficiency.}

\section{System Model}\label{sec:channel}
{Consider} an XL-MIMO HF communication system operating in time division duplexing (TDD) mode, 
where the BS {employs an $N$-element uniform linear array (ULA) to serve single-antenna users}. The HF channel from the BS to the user $\mathbf{h} \in \mathbb{C}^{N \times 1}$ is superposed by the FF path components $\mathbf{h}_F \in \mathbb{C}^{N \times 1}$ ($d > D$) and the NF components $\mathbf{h}_N \in \mathbb{C}^{N \times 1}$ ($d < D$), where $d$ is the distance from the BS to the scatter, $R$ is the aperture of the array antenna, $\lambda$ is the wavelength and $D_{\text{Rayleigh}} = \frac{2R^2}{\lambda}$ is the Rayleigh distance. 
%
%
Then, the FF path components $\mathbf{h}_F$ can be expressed as
\begin{equation}\label{eq:1}
    \mathbf{h}_F = \sqrt{\frac{1}{L}} \sum_{\ell_f=1}^{L_F} \alpha_{\ell_f} \, \mathbf{a}(\theta_{\ell_f}),
\end{equation}
where $\alpha_{\ell_f}$ and $\theta_{\ell_f}$ denote, respectively, the complex gain and the azimuth 
angle of the $\ell_f$-th FF path. From this, the associated array steering vector has the conventional 
form
\begin{equation}\label{eq:2}
    \mathbf{a}(\theta_{\ell_f}) = \left[1, e^{-j\pi \theta_{\ell_f}}, \dots, e^{-j\pi(N-1)\theta_{\ell_f}} \right]^H,
\end{equation}
with $\theta_{\ell_f}=\sin(\theta_{\ell_f}) \in (-\pi/2,\pi/2)$. Since the number of dominant scatterers in the FF is typically limited, $\mathbf{h}_F$ is sparse in the angular domain. {This leads to the representation
\begin{equation}\label{eq:3}
    \mathbf{h}_F = \mathbf{U}\mathbf{h}_f,
\end{equation}
where $\mathbf{U} \in \mathbb{C}^{N \times Q_F}$ is the angular-domain basis. Setting $Q_F = N$ makes $\mathbf{U}$ the $N$-point {\it Discrete Fourier Transform} (DFT) matrix, with $\mathbf{h}_f$ denoting the sparse angle-domain vector.}

On the other hand, the NF components must account for both angular and distance variations. Therefore, they can be modeled as
\begin{equation}\label{eq:4}
    \mathbf{h}_N = \sqrt{\tfrac{1}{L}} \sum_{\ell_n=1}^{L_N} \alpha_{\ell_n}\,\mathbf{a}(\theta_{\ell_n}, r_{\ell_n}),
\end{equation}
where $\alpha_{\ell_n}$, $\theta_{\ell_n}$ and $r_{\ell_n}$ denote the complex gain, azimuth angle 
and distance of the $\ell_n$-th NF path. The corresponding array response is
\begin{equation}\label{eq:5}
    \mathbf{a}(\theta_{\ell_n}, r_{\ell_n}) =
    \left[e^{-j\frac{2\pi}{\lambda}(r_{\ell_n}^{(1)}-r_{\ell_n})}, \dots,
          e^{-j\frac{2\pi}{\lambda}(r_{\ell_n}^{(N)}-r_{\ell_n})}\right]^H,
\end{equation}
where $r_{\ell_n}^{(n)}$ denotes the distance from scatterer $\ell_n$ to the $n$-th antenna element. 
As shown in~\cite{Cui2022}, NF components are also sparse when represented in the polar domain. This sparsity motivates the definition of a polar-domain transformation matrix {$\mathbf{V} \in \mathbb{C}^{N \times Q_N}$}, which allows the compact representation
\begin{equation}\label{eq:7}
    \mathbf{h}_N = \mathbf{V}\mathbf{h}_n,
\end{equation}
with $\mathbf{h}_n$ being sparse in the joint angle–distance domain.

The overall non-line-of-sight (NLoS) contribution is then described by the superposition of FF and NF terms:
\begin{equation}\label{eq:h_nlos}
    \mathbf{h}_{\rm NLOS} = \sqrt{\tfrac{1}{L}}
    \left(\sum_{\ell_f=1}^{\gamma L}\alpha_{\ell_f}\mathbf{a}(\theta_{\ell_f})
         + \sum_{\ell_n=1}^{(1-\gamma)L}\alpha_{\ell_n}\mathbf{a}(\theta_{\ell_n}, r_{\ell_n})\right),
\end{equation}
where $\gamma=L_F/L$ denotes the proportion of FF components. Notice that $L$ is used for channel modeling purposes. The proposed algorithm does not require explicit knowledge of $L$ or $\gamma$ to operate\footnote{For details, see Remark 1}.

Combining LoS and NLoS contributions, the {HF} XL-MIMO channel is finally modeled as a Rician fading process:
\begin{equation}\label{eq:h_los_nlos}
    \mathbf{h} = \sqrt{\tfrac{\kappa}{\kappa+1}}\,\mathbf{h}_{\rm LOS}
               + \sqrt{\tfrac{1}{\kappa+1}}\,\mathbf{h}_{\rm NLOS},
\end{equation}
where $\kappa$ is the Rician factor and $\mathbf{h}_{\rm LOS}=\mathbf{a}(\theta_0,r_0)$ denotes the LoS steering vector, which reduces to the FF form if $r_0$ exceeds the Rayleigh distance. The NLoS gains are modeled as $\alpha_l\sim\mathcal{CN}(0,1/(\kappa+1))$.

For compactness, the HF channel can be rewritten as
\begin{equation}\label{eq:h}
    \mathbf{h} = \mathbf{U}\mathbf{h}_f + \mathbf{V}\mathbf{h}_n + \mathbf{q}_e,
\end{equation}
where $\mathbf{q}_e$ accounts for quantization errors. This representation highlights the dual sparsity of the HF channel in both the angular (DFT) and polar domains, which is the basis for the proposed estimation framework.

From \ref{eq:h}, it is evident that the HF channel can be expressed as a superposition of FF and NF components. Consequently, the estimation of FF and NF channels can be carried out independently.Therefore, we can rewrite \ref{eq:h} as:
\begin{equation}\label{eq:h_eq}
    \mathbf{h} = \mathbf{A} \mathbf{\bar{h}} + \mathbf{q}_e,
\end{equation}
where {$Q = Q_F + Q_N$,} $\mathbf{A} = [ \mathbf{U}, \mathbf{V} ] \in \mathbb{C}^{N \times N + Q}$ is the complete dictionary matrix with both FF and NF sampled points in the angular and polar domains, respectively, while $\mathbf{\overline{h}}\in \mathbb{C}^{N + Q \times 1}$ is the sparse channel vector.

{To estimate the HF channel, the UE transmits an orthogonal pilot sequence over $\tau$ time slots. The received pilot matrix at the BS, $\mathbf{Y} \in \mathbb{C}^{N \times \tau}$, is given by}
\begin{equation}\label{eq:10}
   \mathbf{Y} = \mathbf{h} \boldsymbol{\Phi}^{H} + \mathbf{N},
\end{equation}
where $\boldsymbol{\Phi} \in \mathbb{C}^{\tau \times 1}$ denotes the normalized pilot vector, satisfying $\boldsymbol{\Phi}^{H}\boldsymbol{\Phi} = \tau$ to ensure orthogonality, and $\mathbf{N} \in \mathbb{C}^{N \times \tau}$ is additive noise with i.i.d.\ entries distributed as $\mathcal{CN}(0,\sigma^{2})$.

{To estimate the channel, the BS correlates $\mathbf{Y}$ with the normalized pilot $\boldsymbol{\Phi}/\sqrt{\tau}$, yielding}
\begin{equation}\label{eq:11}
    \mathbf{y}_t = \frac{1}{\sqrt{\tau}} \mathbf{Y} \boldsymbol{\Phi} 
         = \frac{1}{\sqrt{\tau}} \big( \mathbf{h} \, \boldsymbol{\Phi}^{H} \boldsymbol{\Phi} + \mathbf{N} \boldsymbol{\Phi} \big)
         = \sqrt{\tau} \, \mathbf{h} + \mathbf{n}_t ,
\end{equation}
where $\mathbf{n}_t = \tfrac{1}{\sqrt{\tau}} \mathbf{N} {\boldsymbol{\Phi}}\in \mathbb{C}^{N \times 1}$ denotes the equivalent noise after correlation.

\section{Sparse Channel Estimation Formulation}\label{sec:problem}

The {CE} task consists of recovering a sparse vector $\overline{\mathbf{h}}$ from the received signal $\mathbf{y}_t$, ideally by solving
\begin{equation}
\min_{\overline{\mathbf{h}}}\ \|\overline{\mathbf{h}}\|_0 
\quad \text{s.t.}\quad \|\mathbf{y}_t - \mathbf{A}\overline{\mathbf{h}}\|_2 \leq \epsilon,
\end{equation}
where $\mathbf{A}$ is the dictionary and $\epsilon$ limits the residual error. {Because this $\ell_0$ problem is intractable, it is commonly replaced by the residual minimization}
\begin{equation}
\min_{\overline{\mathbf{h}}}\ \|\mathbf{y}_t-\mathbf{A}\overline{\mathbf{h}}\|_2^2,
\end{equation}
or, when the noise variance $\sigma^2$ is known, by its constrained form with $\epsilon=\sigma^2$. {Here, $\epsilon$ controls the allowable mismatch: if too small, the estimator overfits noise; if properly chosen, it enforces a balance between measurement fidelity and robustness, yielding a channel estimate that reflects the true propagation characteristics rather than noise artifacts.}

{\remark {\textbf{\textit{Model Parameters vs. Algorithm Inputs}}. 
The parameter $L$ in \eqref{eq:h_nlos} is used solely to describe the physical channel model and to generate synthetic realizations in simulations. It is not an input to the proposed $\epsilon$-OMP-SSIGW. The estimator does not require $L$ or $\gamma$; instead, it iteratively selects atoms and stops based on the residual criterion $\|\mathbf{r}^{[i]}\|_2^2/N \le \sigma^2$, with $[i]$ being the iteration index, thereby determining the effective number of paths automatically, as detailed in Section~\ref{sec:sol_methods}.}}

\subsection{Proposed Solution and Algorithm}\label{sec:sol_methods}
{To improve HF CE with low computational cost, we adopt a greedy OMP-based approach enhanced by the proposed SSIGW refinement. This allows recovery of channel coefficients in a redundant dictionary spanning both angular and polar domains, capturing FF angular and NF angle–distance information. The proposed $\epsilon$-OMP-SSIGW algorithm iteratively expands the active dictionary, updates path gains through a single-column LS step, reduced to a simple inner product, and refines the corresponding angle and distance of each selected atom. The main steps are summarized below.}

\noindent\rule{.495\textwidth}{0.4pt}
\noindent\textbf{Initialization.} First, we set $\hat{\mathbf A}^{[0]}=\varnothing$ and the residual $\mathbf r^{[0]}=\mathbf y_t$.

\noindent\textbf{(i) Correlation \& selection.} At iteration $i$, we select the atom most correlated with the residual:
\begin{equation}
j^\star=\arg\max_{j}\,\big|[\mathbf A^H \mathbf r^{[i]}]_j\big|,\qquad 
\hat{\mathbf a}^{[i]}\equiv \mathbf A_{(:,j^\star)}.
\end{equation}

{Then, the selected atom is appended to the active set:
\begin{equation}\label{eq:new_A_i}
\hat{\mathbf A}^{[i]}=\big[\hat{\mathbf A}^{[i-1]},\,\,\hat{\mathbf a}^{[i]}\big],
\end{equation}
where $\hat{\mathbf a}^{[i]}$ is added as a new column of the estimated dictionary (with $\hat{\mathbf A}^{[0]}$ initially empty). The corresponding FF/NF parameters are then initialized, \emph{i.e.}, $\hat \theta_{\text{far}}^{[i]}$ or $(\hat \theta_{\text{near}}^{[i]}, \hat r_{\text{near}}^{[i]})$.
}

\noindent\textbf{(ii) Incremental {LS} for gains.} {Let $\hat{\mathbf A}^{[i-1]}$ and $\overline{\mathbf h}^{[i-1]}$ denote the previously selected atoms and their gains. The corresponding residual is then
\[
\mathbf r \triangleq \mathbf y_t - \hat{\mathbf A}^{[i-1]}\,\overline{\mathbf h}^{[i-1]}.
\]
At iteration $i=1$, since no atom has been chosen, $\mathbf r = \mathbf y_t$. The refinement step then reduces to
\begin{eqnarray}\label{eq:min_problem_refinement2}
\min_{\overline h^{[i]},\,\hat\theta^{[i]},\,\hat r^{[i]}} \|\mathbf{y}_t - \hat{\mathbf A}^{[i]}\overline{\mathbf{h}}^{[i]}\|_2^2
 = \min_{\overline h^{[i]},\,\hat\theta^{[i]},\,\hat r^{[i]}}
\| \mathbf r - \hat{\mathbf a}^{[i]}\overline h^{[i]} \|_2^2,
\end{eqnarray}
with $\hat r^{[i]}$ omitted in the FF case. For fixed $(\hat r^{[i]},\hat\theta^{[i]})$, the optimal gain follows the closed-form LS rule
\begin{equation}\label{eq:h_i}
\overline h^{[i]}_{\mathrm{opt}} \stackrel{\|\hat{\mathbf a}^{[i]}\|_2=1}{=}\ 
(\hat{\mathbf a}^{[i]})^H \mathbf r.
\end{equation}}\vspace{-3mm}

\vspace{-2mm}
\noindent{\textbf{(iii) Single-column refinement.} Substituting $\overline h^{[i]}_{\mathrm{opt}}$ into \eqref{eq:min_problem_refinement2} yields
\begin{equation}
\mathcal L(\hat r^{[i]},\hat\theta^{[i]})
= \mathbf r^H\!\left(\mathbf I-\mathbf\Psi^{[i]}\right)\!\mathbf r,
\qquad
\mathbf\Psi^{[i]} \triangleq \hat{\mathbf a}^{[i]}(\hat{\mathbf a}^{[i]})^H.
\end{equation}
As $\mathbf r^H\mathbf r$ is independent of $(\hat r^{[i]},\hat\theta^{[i]})$,
minimizing $\mathcal L$ is equivalent to maximizing the projected energy $\mathbf r^H \mathbf\Psi^{[i]} \mathbf r$. Therefore,
\begin{equation}
\small
\min_{\hat r^{[i]},\,\hat\theta^{[i]}} 
-\,\mathbf r^H \mathbf\Psi^{[i]} \mathbf r
\quad\Longleftrightarrow\quad
\max_{\hat r^{[i]},\,\hat\theta^{[i]}}
\ \mathbf r^H \mathbf\Psi^{[i]} \mathbf r.
\end{equation}}

\vspace{-2mm}
\noindent\textbf{(iv) Gradient updates (Armijo).} {Using the closed-form scalar gradients derived in Appendix~\ref{apx_A}, each parameter is updated via a backtracking Armijo rule,
\begin{align}
\hat\theta^{(n)}&=\hat\theta^{[i]}-\zeta_\theta\,\nabla_{\theta}\mathcal L\big|_{\hat\theta=\hat\theta^{[i]}},\label{eq:up_theta}\\
\text{(NF) }\ \hat\rho^{(n)}&=\hat\rho^{[i]}-\zeta_\rho\,\nabla_{\rho}\mathcal L\big|_{\hat\rho=\hat\rho^{[i]}}\label{eq:up_rho},
\end{align}
where $\rho^{[i]}=1/r^{[i]}$ is uniformly sampled \cite{Cui2022}, and the step sizes $\zeta_\theta,\zeta_\rho$ are set by Armijo backtracking. After each trial step, a refined atom and its gain are recomputed, and the update is accepted only if the residual decreases (Armijo rule). This refinement procedure ends when the step size becomes too small, the gain stabilizes ($\|\mathbf{\overline{h}}^{[i]}(i) - \overline{h}_{\mathrm{prev}}\| < \tau_{\rm th}$, where $\tau_{\rm th}$ is a threshold), or $n = N_{\mathrm{iter}}$.}

\noindent\textbf{(v) Output.} {The previously described procedure stop once $\text{MSE}\le\epsilon$ or $i$ reaches $\mathcal I_{\max}$. The final estimate is computed as $\hat{\mathbf h}=\hat{\mathbf A}^{[i]}\overline{\mathbf h}^{[i]}$, as summarized in Alg.~\ref{alg:eOMP_ssigw}.
\noindent\rule{.49\textwidth}{0.4pt}
}
{Note that our proposed method mitigates grid mismatch via coarse-to-fine refinement, is robust to noise-variance errors, and handles dictionary coherence through orthogonal selection and single-column refinement.}

\section{Numerical Results and Discussions}\label{sec:results}
{In this section, we compare the proposed CE method with existing HF estimators. Simulations assume $\lambda=0.01$ m ($f_c=30$ GHz) and an HF channel with $L=10$ paths and $\gamma=0.5$, parameters used only for Monte–Carlo generation, as the proposed $\epsilon$-OMP-SSIGW operates blindly with respect to both. A pilot length of $\tau{=}1$ is adopted for simplicity, though the method naturally extends to $\tau>1$ since all steps rely on the sufficient statistic $\mathbf{A}^{H}\mathbf{r}$, whose SNR improves linearly with pilot averaging. The stopping rule uses $\epsilon=\sigma^{2}$. In mixed LoS/NLoS scenarios, the LoS component uses $\kappa=10$ (yielding $\bar\alpha_0=\sqrt{10/11}$), while NLoS gains follow $\alpha_\ell \sim \mathcal{CN}(0,1/11)$. User distances are drawn from $\mathcal U(10,500)$ and angles from $\mathcal U(-1,1)$. Additional parameters are summarized in Table~\ref{tab:sim_param}. The NMSE metric is defined as}
\[
\text{NMSE} = \mathbb{E}\left\{\frac{\| \mathbf{\hat{h}} - \mathbf{h} \|^2}{\|\mathbf{h} \|^2}\right\},
\]
where $\mathbf{h} $ denotes the HF channel and $\mathbf{\hat{h}}$ its estimate. 

\begin{table}[!htbp]
\centering
\caption{Simulation Parameters: adopted values}
\label{tab:sim_param}
\centering
\begin{tabular}{cc|cc}
\hline
\textbf{Parameter} & \textbf{Value} & \textbf{Parameter} & \textbf{Value} \\ \hline
$N$                 & 256   & $f_c$                 & 30 GHz\\
$Q_N = Q_F$         & 256   & $Q$                   & 512   \\ $\epsilon$          & $\sigma^2$ &  $\tau_{\rm th}$ & $10^{-1}$ \\ 
$L$                 & 10    & $\gamma$              & 0.5       \\
$\zeta_\theta{=}\zeta_\rho$ & $5\!\times\!10^{-4}$  & $\zeta_{\rm th}$ & $10^{-5}$ \\
$\theta_{\ell_f}$           & $\mathcal{U}[-1,1]$   & $\theta_{\ell_n}$   & $\mathcal{U}[-1,1]$\\ 
$N_{\mathrm{iter}}$         & 5                     & $\mathcal{I}_{\text{max}}$ & 20\\ 
\hline
\end{tabular}
\end{table}

\begin{algorithm}[!hb]
\caption{$\epsilon$-OMP-SSIGW Channel Estimation Scheme.}
\label{alg:eOMP_ssigw}
\small
\begin{algorithmic}[1]
\Statex \textbf{Input:} $\mathbf{y}_{t},\,\mathbf{A},\,\epsilon,\,\mathcal{I}_{\text{max}}$, $N_{\mathrm{iter}}$, $\zeta_\theta$, $\zeta_\rho$, $\zeta_{\rm th}$, $\tau_{\rm th}$
\Statex \textbf{Initialize:} $i \leftarrow 0$, $\mathbf{r} \leftarrow \mathbf{y}_{t}$, $\mathcal{T}^{[i]} \leftarrow \emptyset$, $\hat{\mathbf{A}}^{[i]}\leftarrow \emptyset$, $\overline{\mathbf{h}}^{[i]}\leftarrow \emptyset$ $\text{MSE} \leftarrow 2\epsilon$
\While{$\text{MSE} > \epsilon$ \textbf{and} $i \leq \mathcal{I}_{\text{max}}$}
    \State $i\leftarrow i + 1$
    \State $\eta \leftarrow \text{arg}\max_{j} \| [\mathbf{A}^{H} \mathbf r ]_{j} \|^2$, then  $\hat{\mathbf a}^{[i]} \leftarrow \mathbf{A}_{(:,\eta)}$
    \State $\mathcal{T}^{[i]} \leftarrow \mathcal{T}^{[i-1]}\,\,\cup\,\, \eta$; $\hat{\mathbf{A}}^{[i]} \leftarrow [\hat{\mathbf{A}}^{[i-1]}, \hat{\mathbf a}^{[i]}]$
    \State $\mathbf{\overline{h}}^{[i]} \leftarrow [\mathbf{\overline{h}}^{[i-1]}; (\hat{\mathbf a}^{[i]})^H \mathbf r] $ \eqref{eq:h_i}
    \State Define $\mathbf{r}_{\text{actual}} \leftarrow \mathbf r - \hat{\mathbf{A}}^{[i]}(:,i) \mathbf{\overline{h}}^{[i]}(i)$; $\overline{h}_{\mathrm{prev}} \leftarrow 0$
    \For{\(n=1\) to \(N_{\mathrm{iter}}\)}
        \State compute $\nabla_{\theta}\mathcal L$ (App.~\ref{apx_A}), and define $\zeta \leftarrow \zeta_\theta$
        \While{$\zeta > \zeta_{\rm th}$}
            \State $\theta_{\mathrm{try}} \leftarrow \theta^{[i]} - \zeta\,\nabla_{\theta}\mathcal L$ \eqref{eq:up_theta}    
            \State build $\mathbf a_{\mathrm{try}}$, and evaluate  $\bar h_{\mathrm{try}} \leftarrow \mathbf a_{\mathrm{try}}^{H}\mathbf r$
            \State $\mathbf r_{\mathrm{try}} \leftarrow \mathbf r - \mathbf a_{\mathrm{try}}\,\bar h_{\mathrm{try}}$
            \If{$\|\mathbf r_{\mathrm{try}}\|^{2} < \|\mathbf r_{\mathrm{actual}}\|^{2}$} \Comment{Armijo}
                \State $\hat \theta^{[i]} \leftarrow \theta_{\mathrm{try}}$; $\hat{\mathbf A}^{[i]}(:,i) \leftarrow \mathbf a_{\mathrm{try}}$; $\mathbf{\overline{h}}^{[i]}(i) \leftarrow \overline{h}_{\mathrm{try}}$; $\mathbf r_{\mathrm{actual}} \leftarrow \mathbf r_{\mathrm{try}}$; \Comment{Accept update} 
                \State \textbf{break}
            \Else
                \State $\zeta \leftarrow \zeta/2$
            \EndIf
        \EndWhile
        \If{NF case}
            \State repeat the same refinement procedure for $\rho^{[i]}$ 
        \EndIf
        \If{$\|\mathbf{\overline{h}}^{[i]}(i) - \overline{h}_{\mathrm{prev}}\| < \tau_{\rm th}$} \Comment{Check gain convergence}
        \State \textbf{break}
    \Else
        \State $\overline{h}_{\mathrm{prev}} \leftarrow  \mathbf{\overline{h}}^{[i]}(i)$
    \EndIf
    \EndFor
    \State Update $\mathbf r \leftarrow \mathbf r - \hat{\mathbf{A}}^{[i]}(:,i)\mathbf{\overline{h}}^{[i]}(i)$ 
    \State $\text{MSE} = \frac{1} {N}\text{Tr}\left[\mathbf r^H\mathbf r\right]$    
\EndWhile
\Statex \textbf{Output:} $\mathbf{\hat{h}} = \hat{\mathbf{A}}^{[i]} \mathbf{\overline{h}}^{[i]}$ 
\end{algorithmic}
\end{algorithm}

\begin{figure}[!htbp]
 \centering
\includegraphics[width=0.45\textwidth]{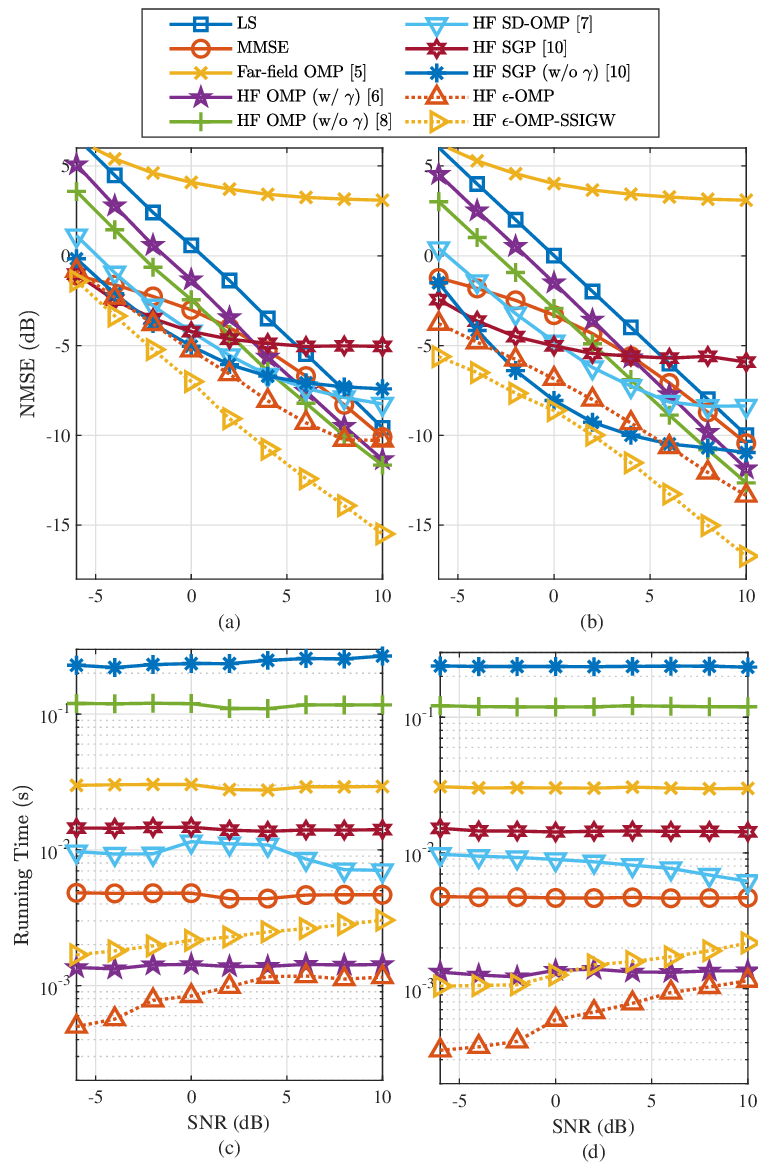}
\vspace{-3mm}
\caption{{NMSE and runtime of LS, MMSE, FF-OMP, HF-OMP variants, SD-OMP, HF SGP with $\gamma$, HF SGP without $\gamma$, and the proposed $\epsilon$-OMP-SSIGW for (a) NLoS, (b) mixed LoS/NLoS, and (c)–(d) runtime results.}}
\label{fig:NMSEvsSNR_NLOS}
\end{figure}

{Fig.~\ref{fig:NMSEvsSNR_NLOS} compares the NMSE of the evaluated CE schemes under NLOS ($\kappa{=}0$) and mixed LOS/NLOS ($\kappa{=}10$) conditions. In Fig.~\ref{fig:NMSEvsSNR_NLOS}(a), LS exhibits poor performance across all SNRs, while MMSE performs well at low SNR but rapidly degrades as SNR increases. All HF-based schemes outperform these baselines, mainly at high SNRs, with the proposed $\epsilon$-OMP achieving the lowest NMSE over nearly the entire SNR range. Incorporating SSIGW refinement yields further gains and provides the best overall accuracy even under pure NLOS conditions. In the mixed LOS/NLOS scenario of Fig.~\ref{fig:NMSEvsSNR_NLOS}(b), HF-SGP, SGP+SIGW, and HF-OMP variants partially mitigate grid mismatch but still fall short, with HF-SGP exhibiting a visible error floor. The proposed $\epsilon$-OMP-SSIGW consistently attains the best NMSE, demonstrating strong robustness across propagation conditions.
}
\vspace{-2mm}

{The runtime curves in Fig.~\ref{fig:NMSEvsSNR_NLOS}(c)–(d) indicate SNR-independent complexity, dominated by fixed matrix operations. LS is the fastest ($10^{-5}$–$10^{-4}$~s) and is omitted for clarity. Among HF schemes, $\epsilon$-OMP is the fastest, while the refined $\epsilon$-OMP--SSIGW incurs only minor overhead and delivers substantial NMSE gains. HF-OMP with $\gamma$ is slightly faster but less accurate, whereas HF-OMP (without $\gamma$) and HF-SGP exhibit significantly higher runtimes, confirming the favorable performance–complexity trade-off of the proposed method.}

\begin{figure}[!htbp]
 \centering
\includegraphics[width=0.47\textwidth]{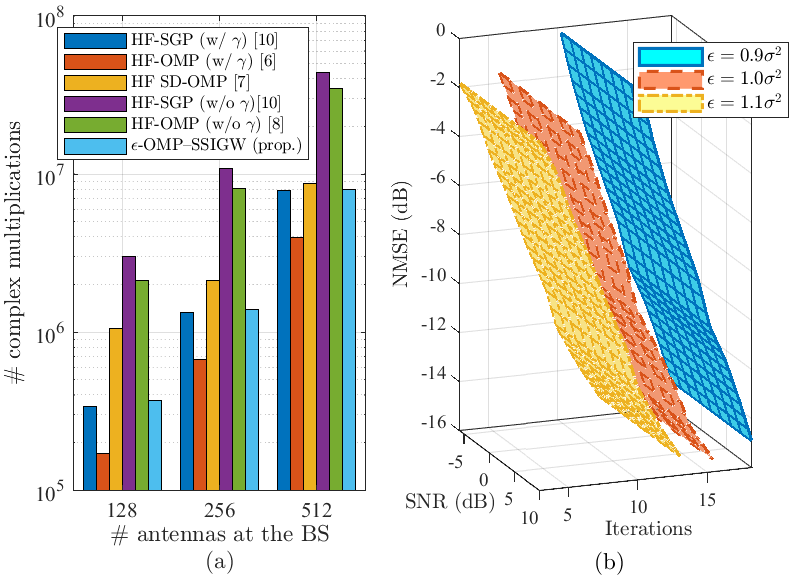}
\vspace{-3mm}
\caption{{(a) Complexity results based on Table~\ref{tab:complexity_updated};
             (b) NMSE and the resulting $L_{\rm est}$ (mean iteration count of the $\epsilon$-OMP-SSIGW) for different values of~$\epsilon$.}} 
\label{fig:complexity_and_nmse}
\end{figure}

{The computational complexity in Table~\ref{tab:complexity_updated} reflects the dominant per-iteration operations of each HF estimator. For the proposed $\epsilon$-OMP--SSIGW, the cost is governed by the correlation step $\mathbf{A}^H\mathbf{r}$, the incremental LS gain update, the vector-level SSIGW refinement, and the final $\mathcal{O}(Ni)$ reconstruction. This yields a total complexity of $\mathcal{O}(NQi + NN_{\mathrm{iter}}Bi + Ni)$, consistent with prior HF analyses. Here, $B$ is the number of Armijo backtracking attempts. The results in Fig.~\ref{fig:complexity_and_nmse}, under the parameters of Table~\ref{tab:sim_param} unless otherwise stated, show that the effective number of paths ($L_{\rm est}=$ Iterations) varies with SNR and differs from the true $L$, confirming that the estimator does not require $L$ as input. The NMSE surfaces further demonstrate robustness to noise-variance mismatch, with at most a 1~dB loss for $\epsilon=0.9\sigma^2$ and even less with $1.1\sigma^2$. Combined with its linear-in-$i$ scaling, these results highlight the favorable accuracy–complexity trade-off of the proposed method.}

\vspace{-2mm}
\begin{table}[!htbp]
\centering
\caption{{Computational complexity of HF CE schemes.}} 
\label{tab:complexity_updated}
\renewcommand{\arraystretch}{1.5}
\resizebox{0.49\textwidth}{!}{\begin{tabular}{l|c}
\hline
\textbf{Scheme} & \textbf{Complexity} \\
\hline
HF OMP w/ $\gamma$ \cite{Wei2022} & 
$\mathcal{O}\!\big(N(Q_F L_F + L_F^2) + L_F^3 + N(Q_N L_N + L_N^2) + L_N^3\big)$ \\
HF OMP w/o $\gamma$ \cite{Yang2023} &
$\mathcal{O}\!\big(N Q_F L + N L^2 + L^3 + N_{\gamma}(N Q_N + N L^2 + L^3 + N^2 L)\big)\ $ \\
HF SD-OMP \cite{Hu2023} & $\mathcal{O}\!\big(N L_F O^2 + (L_F O)^2 + Q_N(L_N O)^3 + N Q_N \big)$ \\
HF SGP w/ $\gamma$ \cite{Lei2024} &
$\mathcal{O}\!\big(NL^2 + N L (Q_F + Q_N) \big)$ \\
HF SGP w/o $\gamma$ \cite{Lei2024} &
$\mathcal{O}\!\big(N Q_F L + N L^2 + N_{\gamma}(N Q_N + N L^2)+ N_{\text{iter}}(N^2 + N L^2 + L^3 + N^2 L)\big)$ \\
{\textbf{$\epsilon$-OMP-SSIGW}} &
{$\mathcal{O}\!\big(iN(Q + BN_{\mathrm{iter}} + 1)\big)$}\\
\hline
\end{tabular}}
\end{table}
\vspace{-4mm}

\vspace{-5mm}
\section{Conclusions}\label{sec:conclusions}
{We proposed an HF CE framework for XL-MIMO systems, combining an $\epsilon$-OMP selection stage with the derived SSIGW refinement. Numerical results show that the proposed estimator achieves the lowest NMSE among all benchmark methods, outperforming LS/MMSE and all HF-OMP and HF-SGP variants under both LoS and NLoS conditions.}

{Regarding complexity, Figs.~\ref{fig:NMSEvsSNR_NLOS}(c)–(d) show that the proposed $\epsilon$-OMP-SSIGW is exceeded only by LS and its own non-refined variant, while delivering substantially higher CE accuracy than all of the considered baselines. This efficiency stems from incremental LS gain updates that avoid repeated pseudoinverses and from a lightweight scalar refinement stage in which FF/NF gradients reduce to inner products rather than matrix operations, as in \cite{Cui2022,Lei2024}. Consequently, the method achieves a favorable accuracy–complexity balance, as further illustrated in Table~\ref{tab:complexity_updated} and Fig.~\ref{fig:complexity_and_nmse}.}

{Finally, natural extensions include: multi-antenna UE scenarios, generalization to UPA/XL-array geometries, incorporation of hardware impairments, and exploration of more advanced dictionary-learning or data-driven refinement strategies.}

\vspace{-5mm}

\appendices

\section{Gradient 
for Single-Column Refinement}\label{apx_A}

We derive the gradient of the cost function in \eqref{eq:min_problem_refinement2} with respect to any parameter $u$ associated with $\hat{\mathbf{a}}^{[i]}$, given by:

\begin{equation}\label{eq:dLdz_start}
{\frac{\partial \mathcal L}{\partial u}
= -\mathbf{r}^H \frac{\partial\mathbf\Psi}{\partial u} \mathbf{r}.}
\end{equation}

From the development of \eqref{eq:dLdz_start}, we obtain
\begin{equation}
{\frac{\partial \mathcal L}{\partial u}
= -2\,\Re\!\big\{ z^* \, (\partial_u\mathbf a)^H \mathbf r \big\}, 
\quad z \triangleq \mathbf a^H \mathbf r.}
\label{eq:dLdu_compact_short}
\end{equation}

Thus, the computation reduces to two scalar products: {$z=\mathbf a^H\mathbf r$ and $w=(\partial_u\mathbf a)^H\mathbf r$}, yielding $\partial \mathcal L/\partial u = -2\Re\{z^* w\}$.

\noindent\textbf{Far-field (FF) case:} $\mathbf{a}^{[i]}$ depends only on the angle $\theta_i$, with element-wise derivative
\begin{equation}
\left[\frac{\partial \mathbf{a}^{[i]}}{\partial \theta_i}\right]_n 
= -j \pi (n-1) \left[\mathbf{a}^{[i]}\right]_n.
\end{equation}

\noindent\textbf{Near-field (NF) case:} $\mathbf{a}^{[i]}$ depends on both angle $\theta_i$ and inverse distance $\rho_i=1/r_i$. Approximating the spherical wave phase to second order,
\begin{equation}
\left[\mathbf{a}^{[i]}\right]_n = e^{-j\frac{2 \pi f}{c} r_m}, 
\quad r_m \approx - m d\,\theta_i + \tfrac{(md)^2}{2}(1-\theta_i^2)\rho_i,
\end{equation}
with $m=(n-1)-(N-1)/2$. The derivatives are
\begin{align}
\left[\tfrac{\partial \mathbf{a}^{[i]}}{\partial \theta_i}\right]_n 
&= - j \tfrac{2 \pi f}{c} \big( - m d - (md)^2 \theta_i \rho_i \big) \left[\mathbf{a}^{[i]}\right]_n, \\
\left[\tfrac{\partial \mathbf{a}^{[i]}}{\partial \rho_i}\right]_n 
&= - j \tfrac{\pi f}{c}(md)^2(1-\theta_i^2) \left[\mathbf{a}^{[i]}\right]_n.
\end{align}

These gradients, {$\nabla_{\theta_i}\mathcal L$} and {$\nabla_{\rho_i}\mathcal L$}, are then applied in the iterative refinement of the chosen column parameters. {After each accepted update, a refined atom is reconstructed using the new $(\theta_i,\rho_i)$ values and replaces the previous one in the active dictionary matrix, enabling progressive improvement of the estimate.}
\vspace{-3mm}



\end{document}